\documentclass[10pt,letterpaper]{article}
\usepackage[top=0.85in,left=2.75in,footskip=0.75in,marginparwidth=2in]{geometry}

\usepackage[utf8]{inputenc}

\usepackage{cite}

\usepackage{nameref,hyperref}

\usepackage[right]{lineno}

\usepackage{microtype}
\DisableLigatures[f]{encoding = *, family = * }

\raggedright
\usepackage{changepage}

\usepackage[aboveskip=1pt,labelfont=bf,labelsep=period,singlelinecheck=off]{caption}

\makeatletter
\renewcommand{\@biblabel}[1]{\quad#1.}
\makeatother

\usepackage{lastpage,fancyhdr,graphicx}
\usepackage{epstopdf}
\fancyheadoffset[L]{2.25in}
\fancyfootoffset[L]{2.25in}

\usepackage{color}

\definecolor{Gray}{gray}{.25}

\usepackage{graphicx}

\usepackage{sidecap}

\usepackage{wrapfig}
\usepackage[pscoord]{eso-pic}
\usepackage[fulladjust]{marginnote}
\reversemarginpar

\begin{document}
\vspace*{0.35in}

\begin{flushleft}
{\Large
\textbf\newline{How fear of future outcomes affects social dynamics}
}
\newline
\\
Boris Podobnik\textsuperscript{1,2,3,*},
Marko Jusup\textsuperscript{4},
Zhen Wang\textsuperscript{5},
H. Eugene Stanley\textsuperscript{1},
\\
\bigskip
\bf{1} Center for Polymer Studies and Department of Physics, Boston University, Boston, MA 02215
\\
\bf{2} Faculty of Civil Engineering, University of Rijeka, 51000 Rijeka, Croatia
\\
\bf{3} Zagreb School of Economics and Management, 10000 Zagreb, Croatia
\\
\bf{4} Center of Mathematics for Social Creativity, Hokkaido University, Sapporo 060-0812, Japan
\\
\bf{5} Interdisciplinary Graduate School of Engineering Sciences, Kyushu University, Fukuoka 816-8580, Japan
\\
\bigskip
* bp@phy.hr

\end{flushleft}

\section*{Abstract}
Mutualistic relationships among the different species are ubiquitous in
nature. To prevent mutualism from slipping into antagonism, a host often
invokes a ``carrot and stick'' approach towards symbionts with a stabilizing
effect on their symbiosis. In open human societies, a mutualistic relationship
arises when a native insider population attracts outsiders with benevolent
incentives in hope that the additional labor will improve the standard of all.
A lingering question, however, is the extent to which insiders are willing to
tolerate outsiders before mutualism slips into antagonism. To test the
assertion by Karl Popper that unlimited tolerance leads to the demise of
tolerance, we model a society under a growing incursion from the outside.
Guided by their traditions of maintaining the social fabric and prizing
tolerance, the insiders reduce their benevolence toward the growing
subpopulation of outsiders but do not invoke punishment. This reduction
of benevolence intensifies as less tolerant insiders (e.g., ``radicals'')
openly renounce benevolence. Although more tolerant insiders maintain
some level of benevolence, they may also tacitly support radicals out
of fear for the future. If radicals and their tacit supporters achieve
a critical majority, herd behavior ensues and the relation between the
insider and outsider subpopulations turns antagonistic. To control the
risk of unwanted social dynamics, we map the parameter space within which
the tolerance of insiders is in balance with the assimilation of outsiders,
the tolerant insiders maintain a sustainable majority, and any reduction
in benevolence occurs smoothly. We also identify the circumstances that
cause the relations between insiders and outsiders to collapse or that
lead to the dominance of the outsiders.

\paragraph*{Keywords:} game theory | complex networks | social thermodynamics | open systems  | tolerance | herd behavior

\newpage


\section*{Introduction}
Karl Popper famously stated that unlimited tolerance leads to the demise
of tolerance \cite{Popper}. The tag-based \cite{Axelrod84} quantitative
model of Riolo, Cohen, and Axelrod \cite{Axelrod01} indicates that a
combination of kin selection and mutation causes times of high tolerance
to be replaced by times of low tolerance towards those who are
different. These tides of (in)tolerance \cite{Nowak01} seemingly dismiss
the role of human reasoning in the selection process as unable to stave
off periods during which undesirable states of affairs prevail. In
contrast, the basic negative result of evolutionary game theory that
unconditional cooperators are vulnerable to rare occurrences of
unconditional defectors \cite{Sigmund10} can be avoided by appending an
indirect reciprocity mechanism to the selection process, i.e., a concern
for such abstract realities as reputation, and the ability to use some
form of language to spread information. It would thus seem that humans
\emph{are} able to draw on such collective mechanisms as democracy
\cite{Nowak14} to adjust the course of selection in a preferable
direction---and sometimes they are not \cite{Jusup14}.

Humans are conditional cooperators \cite{Axelrod84, Axelrod81, Nowak08},
yet on occasion may benevolently help even those who can never repay in
kind. Although this cooperativeness is seen as the result of evolutionary
selection \cite{Axelrod84, Axelrod01, Nowak14, Axelrod81, Nowak08, NowakMay92,
Nowak98, Nowak06, Oliviera}, it is less clear why benevolence would permeate
human societies. Popper's statement, however, helps us identify how certain
mechanisms may sustain benevolence. When a benevolent population is
attracting an inflow from the outside, such that a society undergoes the
transient dynamics, understanding the relationship between the inflow rate and
human behavior is key. If the rate of inflow is low the original population
may feel safe, but if the inflow is high it may be perceived as aggression and
provoke---in a sort of Popperian twist---a violent response. The very idea of
these two limits suggests that benevolence is a relative category and is
dependent on the inflow from the outside and the resulting state \cite{Huang15}.
Little is known about these dependencies of benevolence.

\section*{Model}
To quantify the dynamics of an open society with a benevolent population,
we create a theoretical framework by combining the elements of biology (e.g., evolutionary games) and statistical physics (e.g., complex networks). It was argued almost a decade ago that the methods of statistical physics could contribute to social science \cite{Durlauf99}. This notion of the usefulness of statistical physics in researching social phenomena appears to have been appreciated in quantitative sociology, particularly the works on economic complexity \cite{Hidalgo07, Tacchella13}. In line with these works, we represent human relationships in an idealized manner by placing agents into a random network of friendships wherein each possible link occurs independently with precisely such a probability that the average degree is 50---a number consistent with Refs. \cite{Dunbar92, Hill03}.
Agents in the model are thus conditional cooperators in the sense
that their interactions are restricted only to the nearest
neighbors defined by the network. At each time $t$, a total of $m$
donor-recipient pairs are randomly chosen \cite{Nowak98} among neighboring
agents, whereupon a donor pays cost $c$ for the recipient to receive benefit
$b$. To this trivial scenario we add an asymmetry in which insiders incur a
higher cost of cooperation and provide more benefit to outsiders than they
receive in return. In this way, benefit differential $\Delta b$ emerges
between insider and outsider subpopulations, creating an incentive for
outsiders to immigrate. The overall purpose is to form a mutualistic relationship
in which everyone experiences a higher standard due to the extra labor provided
by outsiders. When the selected $m$ donor-recipient pairs finish interacting,
we calculate the fitness of both insiders and outsiders---denoted $\Phi_1$ and
$\Phi_2$, respectively---as the average per-capita benefit net of the cost of
cooperation. The details on the mathematical representation of the described
setup are found in Supporting Information (SI text). Here, we just note that
the quantities of interest are the cost-benefit ratio, $c/b$, and the relative
benefit differential, $\Delta_b/b$. A list of key symbols with some default
parameter values is given in Table~\ref{t1}.

We envision a society that is an open dynamic system in the sense that
its size changes over time. To that end we adopt replicator-type equations.
Specifically, if $N_1$ and $N_2$ denote the population sizes of insiders and
outsiders, respectively, then the time-change of these two subpopulations is
given by $N_i(t+1)=\Phi_i N_i(t)$, $i=1,2$. The fraction of outsiders in the
system is defined as $f_g=N_2/(N_1+N_2)$, yielding
\begin{equation}
  f_g(t+1)=\frac{\Phi_2(t) N_2(t)}{\Phi_1(t) N_1(t) + \Phi_2(t) N_2(t)} =
  R(t) f_g(t),
\end{equation}
where $R$ is the ratio of the fitness of outsiders to the average fitness of
the whole population, i.e., $R=\Phi_2/\left[ \Phi_1 (1-f_g) + \Phi_2 f_g
\right]$. Here the time-change is both biological and sociological, i.e.,
successful individuals can attract migrants of similar origin from the
outside.

We assume that outsiders can ``mutate'' into insiders (i.e., they can
assimilate the cultural patterns of the insiders) and vice versa (e.g.,
due to education or intermarriage). Assimilation, often almost synonymously referred to as integration, has been receiving considerable attention in the literature based on statistical physics \cite{Barra14, Agliari14, Agliari15, Barra16}.
Herein, we represent the results of the assimilation process by a net
rate $p_1$. The net rate is either positive or negative, where positive values
indicate that the absolute assimilation rate is higher for outsiders than insiders.
Our analysis, however, revolves only around the case $p_1>0$ because, as
long as the asymmetry in the system persists, and thus $R>1$,
outsiders grow at a faster rate than insiders. In this sense, fitness ratio
$R$ is a global measure of benevolence exhibited by insiders towards
outsiders. For the two subpopulations to equilibrate, the necessary
and sufficient condition is $p_1=R-1$, which indicates that the more
benevolent insiders are, the higher the net assimilation rate needs to be
for insiders to avoid being overrun by outsiders. If we allowed $p_1<0$,
the two subpopulations could equilibrate only at some $R < 1$, which is
impossible unless insiders start to punish outsiders.

\begin{wrapfigure}{l}{70mm}
\centering \includegraphics[scale=0.77]{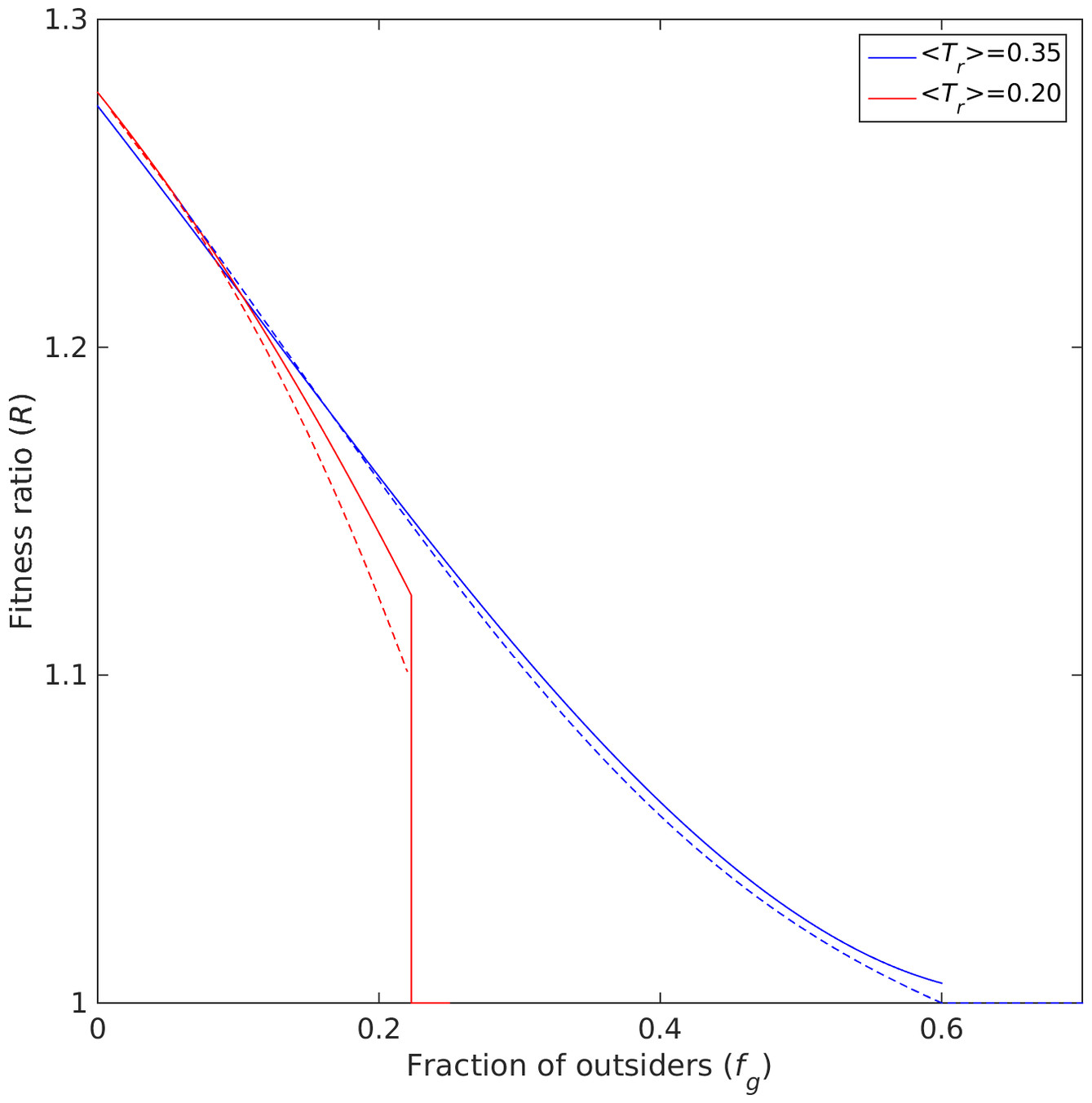} \\
\captionsetup{labelformat=empty} 
\caption{} 
\label{1} 
\end{wrapfigure}

\marginpar{
\vspace{.7cm} 
\color{Gray} 
\textbf{Fig~\ref{1}. Benevolence is dependent on the ubiquity of outsiders.} 
  Benevolence, as reflected by fitness ratio $R$, is a decreasing
  function of the fraction of outsiders, $f_g$, although no explicit
  assumptions were made to that effect. If the average tolerance of
  insiders is low enough (red curve), herd behavior causes a discontinuity
  of function $R=R(f_g)$. Otherwise (blue curve), the functional dependence
  is continuous. Dashed curves are analytical approximations (SI text).
  The assimilation rate is $p_1=0.01$.
}

How do insiders decide whether to cooperate with outsiders? Relying on
the network of friendships as another layer of social fabric, insiders
face three options: (i) remain benevolent, (ii) behave benevolently,
but question the wisdom of doing so, or (iii) no longer cooperate with
outsiders. Insiders who are surrounded by an overwhelming number of
outsiders can begin to resent providing benefit without receiving the
same in return. Thus when the local fraction of outsiders, $f_{\ell}^i$,
in the neighborhood of insider $i$ exceeds the tolerance threshold
\cite{Schelling71, Watts01, Lee15}, $T_r^i$, the insider defects by severing
all connections with outsiders. Insiders who terminate all their friendships
with outsiders we label ``radical agents.'' As a proxy for $T_r^i$ we
take a uniform random variable on the interval from $T_r^{\rm min}$ to
$T_r^{\rm min}+\sigma$, i.e., $T_r^i \sim \mathcal{U}(T_r^{\rm min},
T_r^{\rm min}+\sigma)$. The higher the value of $T_r^{\rm min}$, the greater
the willingness of insiders to tolerate outsiders. Parameter $\sigma$ is a
measure of the degree of individuality, while its inverse $1/\sigma$ measures
societal responsiveness---the lower its value, the greater the extent to which
insiders are willing to tolerate a large benefit differential. When
$\sigma=0$, individuality among the insiders disappears and threshold
distribution $\mathcal{U}$ degenerates into a Dirac delta distribution.
However, $\sigma \ne 0$ is more realistic because tolerance is likely to vary among
people. It is important that the complex network of friendships, depending on the choice of tolerance parameters $T_r^{\rm min}$ and $\sigma$ relative to assimilation parameter $p_1$, may generate herd behavior \cite{Helbing12} such that most connections between insiders and outsiders get abruptly broken \cite{Lee15}.

\begin{table}[!t]
  \centering
  \caption{\small {\bf Key symbols (parameters and state variables)}}
  \begin{tabular*}{\hsize}{@{\extracolsep{\fill}}llc}
    Symbol                & Definition                                      & Value \cr
    \hline
    $c/b$                 & cost-benefit ratio                              & 10\% \cr 
    $\Delta_b/b$          & relative benefit differential                   & 25\% \cr 
    $p_1$                 & assimilation rate                               & --   \cr
    $T_r^\mathrm{min}$    & minimum tolerance of insiders                   & 0.05 \cr
    $\sigma$              & tolerance range (individuality)                 & --   \cr 
    $I_c$                 & critical threshold for herd behavior            & 50\% \cr 
    \hline
    $f_g$                 & \multicolumn{2}{l}{fraction of outsiders in the total population}   \cr
    $R$                   & \multicolumn{2}{l}{fitness ratio measuring benevolence of insiders} \cr
    $\langle T_r \rangle$ & \multicolumn{2}{l}{average population tolerance}                    \cr
    \hline
  \end{tabular*}
  \label{t1}
\end{table}

\begin{figure}[!b]
  \centering \includegraphics[scale=0.77]{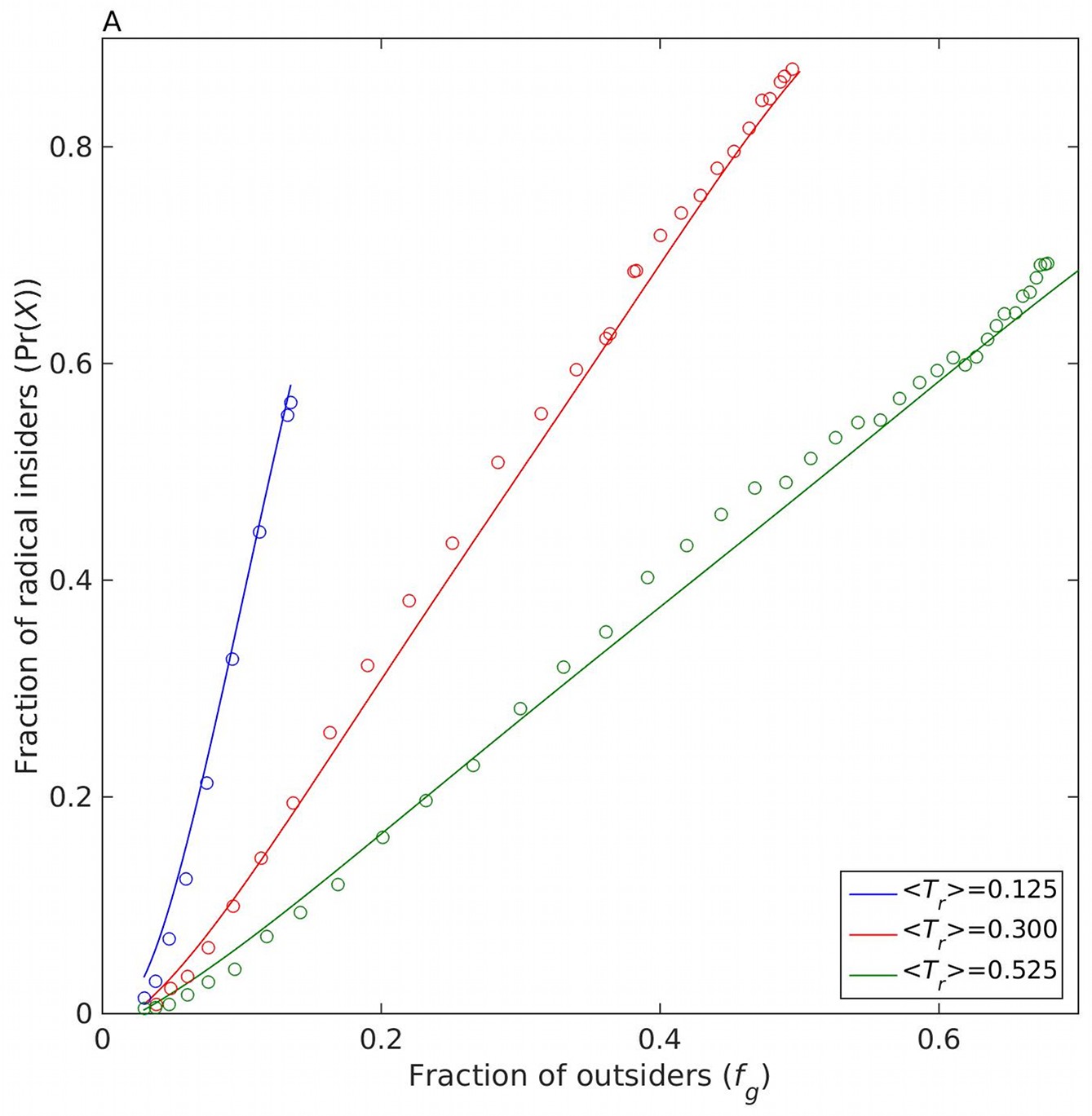}
  \centering \includegraphics[scale=0.77]{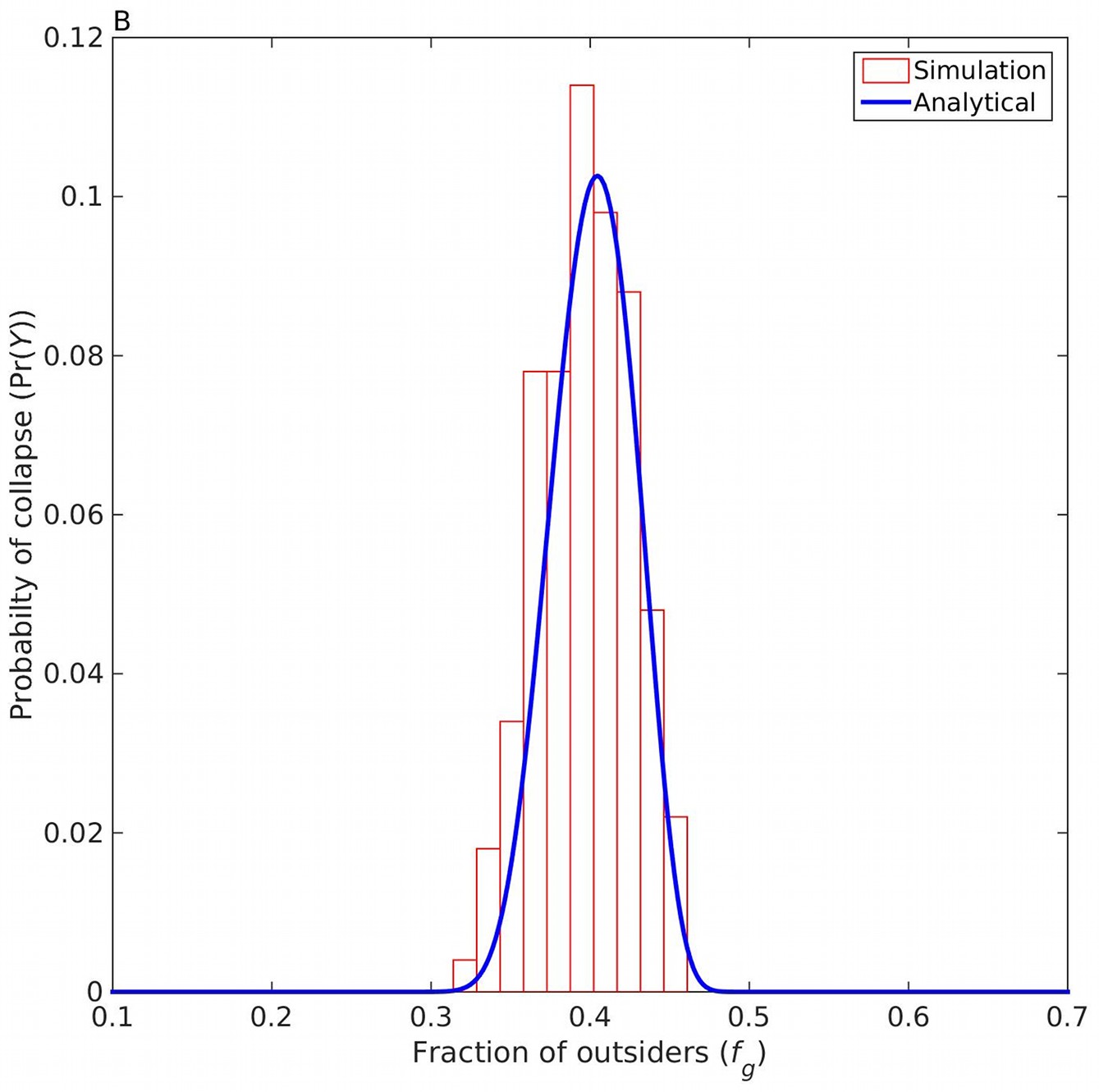}\\
  \caption{{\bf Mean-field approximation of the implied social
      dynamics}. {\bf A}, The fraction of radical insiders is a
    monotonically increasing function of the fraction of outsiders until
    the latter reach a saturation level. Circles indicate the simulation results.
    {\bf B}, The probability of herd behavior during which cooperation
    between insiders and outsiders collapses need not always increase
    with the fraction of outsiders because a high level of outsiders may
    preclude radicals and their tacit supporters from reaching the necessary
    majority. Here we study the case in which $I_c$ is set to 50\% of the total
    population. The assimilation rate is $p_1=0.01$.}
  \label{2}
\end{figure}

Envisioning herd behavior in a different yet related context
\cite{Helbing12}, as the fraction of outsiders ($f_g$) increases, the
fraction of radical agents increases, and the probability of violent
incidents occurring also increases.  As violence increases in a society,
tolerant insiders ($T_r^i \gg f_g$) may consider open animosity towards
outsider friends overly radical, yet adopt a conservative position
\cite{Makse15} that, in effect, supports the radical faction---option
(ii) above. There emerges a disparity between how some insiders act and
think \cite{Hoffman15}. If $g_t$ is the fraction of tacit supporters of
radicals in a large population of $N_1$ insiders, then the probability
of event $Y=\lbrace N_1 g_t \geq I_c \rbrace$ is approximated by a
cumulative Poisson distribution, i.e., $\Pr(Y)=\exp(-\lambda) \sum_{k
  \geq I_c} \lambda^k / k!$, where $\lambda$ is a size-dependent
variable that increases with the global fraction of outsiders $f_g$.
Specifically, we set $\lambda = N_1 (1 - g_r) f_g$, where $g_r$ is the
fraction of insider radical agents. The term in parentheses indicates
that radicals cannot be tacit supporters at the same time.  We further
assume that the increase in tacit supporters causes tension that, at
some critical $I_c$, becomes a violent reaction. Herd behavior ensues,
resulting in the abrupt and simultaneous severing of all connections
between insiders and outsiders. An example of a critical $I_c$ in a
democracy would be a value that gives a majority to radicals and their
supporters during an election. Fig.~\ref{1} shows the basic aspects of the
implied social dynamic, i.e., benevolence as a function of the fraction
of outsiders in the system.

Mean-field theory provides a deeper understanding of the social dynamics
implied by the model. For example, we derive the probability that
randomly picked insider $i$ is radical, i.e., the probability of event
$X=\lbrace f_{\ell}^i \geq T_r^i \rbrace$, where $f_{\ell}^i$ is the
local fraction of outsider neighbors surrounding agent $i$ and $T_r^i$
is $i$'s tolerance. In the mean field approximation, a substitute for
$f_{\ell}^i$ is its global counterpart, $f_g$, which can further be
multiplied with the average degree of the network, $\langle k \rangle$,
to estimate the number of outsider neighbors of an insider. The insider
can tolerate any number of outsiders between $k_{\rm min}=\lceil
T_r^{\rm min} \langle k \rangle \rceil$ and $k_{\mathrm{max}}=\lceil
(T_r^{\rm min}+\sigma) \langle k \rangle \rceil$ with equal probability,
where $\lceil \cdot \rceil$ is the ceiling function. Consequently we
obtain
\begin{equation}
  \Pr(X)=\frac{1}{k_{\mathrm{max}}-k_{\rm min}+1} \sum\limits_{l=k_{\rm
      min}}^{k_{\mathrm{max}}} \sum\limits_{k=l}^{\langle k \rangle}
     {\langle k \rangle \choose k} f_g^k (1-f_g)^{\langle k \rangle -
       k}.
  \label{eq1}
\end{equation}
Fig.~\ref{2}A compares the results of Eq.~(\ref{eq1}) with the corresponding
numerical simulations. Note that a population of insiders with a higher
than average tolerance contains fewer radicals for a given fraction of
outsiders and allows outsiders to reach a higher saturation level. The
agreement between the analytical results and the simulations is
favorable.

Using Eq.~(\ref{eq1}) we can characterize the dependence of the
probability of herd behavior on the fraction of outsiders. In accordance
with our definitions, herd behavior occurs when radicals and their tacit
supporters reach a critical threshold $I_c$. Because the number of
radicals in the system is given by $N_1 g_r$, we only need to consider
the probability of event $Y=\lbrace N_1 g_t \geq I_c - N_1 g_r \rbrace$,
which---as described above---is given by the cumulative Poisson
distribution. In general, the population of insiders, $N_1$, is large and
parameter $\lambda \propto N_1$ of the Poisson distribution is also large,
and this allows us to use the normal approximation for probability
$\Pr(Y)$. Accordingly, $\Pr(Y) \approx 1 - F_{\mathrm{norm}}(I_c/N_1-g_r;
\lambda/N_1, \sqrt{\lambda}/N_1)$, where $F_{\mathrm{norm}}$ is the
cumulative distribution function of a normal random variable with mean
$\lambda/N_1 = (1-g_r) f_g$ and standard deviation $\sqrt{\lambda}/N_1 =
\sqrt{(1-g_r) f_g/N_1}$. During the calculations $g_r$ is replaced with
$\Pr(X)$ from Eq.~(\ref{eq1}). 
Fig.~\ref{2}B shows the favorable comparison
between this approximation and the simulation results and indicates that
herd behavior becomes possible only when the fraction of outsiders is
already high, i.e., $f_g \approx 0.4$. As $f_g$ continues to increase,
the possibility of herd behavior becomes increasingly remote because it
is now difficult for radicals and their tacit supporters to acquire an
electoral majority.

\begin{wrapfigure}{l}{70mm}
\centering \includegraphics[scale=0.77]{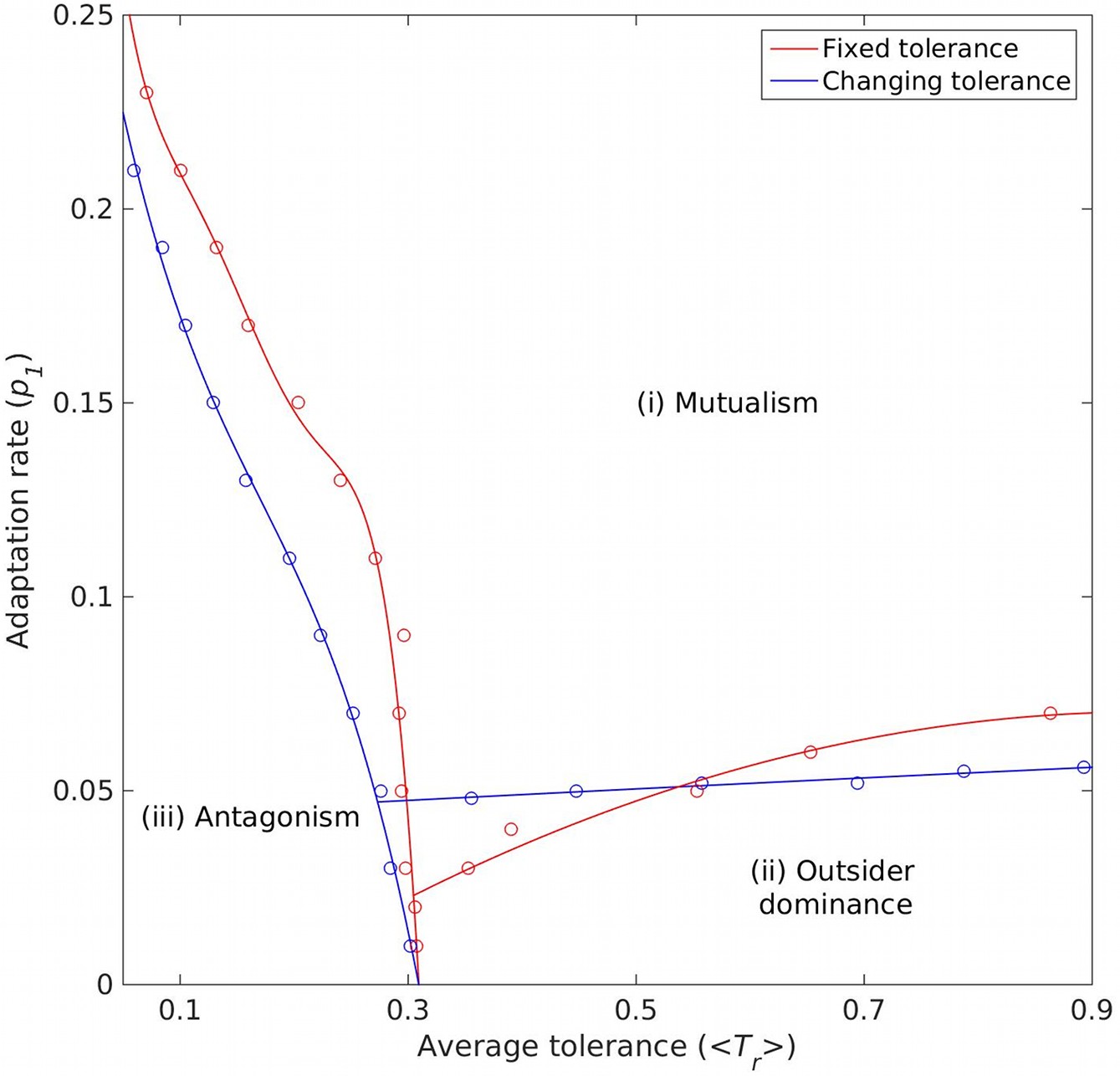}\\
\captionsetup{labelformat=empty} 
\caption{} 
\label{3} 
\end{wrapfigure}

Although analytical results help us understand the dynamics of the
model, comprehensive mapping of the parameter space is often impossible
without resorting to numerical simulations. Before detailing the results of these simulations, we emphasize how
Fig.~\ref{1} shows that benevolence measured by fitness ratio $R$ is
a decreasing function of the fraction of outsiders $f_g$. A diminishing
benevolence in the presence of a benefit differential is hardly
surprising given that benevolent populations disappear by virtue of
evolutionary dynamics. The question, therefore, is whether and under
what conditions adjusting benevolence is sufficient to accommodate an
incursion of outsiders without causing societal turmoil. 
Accordingly, we recognize three different types of outcomes in numerical simulations (Fig.~\ref{3}):
\begin{description}
\item[\textit{Mutualism}] is a set of equilibrium states reached by a smooth reduction of benevolence to a level at which insiders maintain a sustainable majority (continuous blue curve in Fig.~\ref{1}).
\item[\textit{Outsider dominance}] is a set of equilibrium states wherein the outsiders form a majority.
\item[\textit{Antagonism}] is a set of non-equilibrium, absorbing states due to a complete breakdown of cooperation between the two subpopulations (discontinuous red curve in Fig.~\ref{1})
\end{description}
\noindent We find that over a considerable portion of the parameter space the tolerance of
insiders is in balance with the assimilation of outsiders, and there is
a smooth reduction of benevolence to a level at which insiders maintain
a sustainable majority (see the \emph{Mutualism} region in Fig.~\ref{3}).

\marginpar{
\vspace{.7cm} 
\color{Gray} 
\textbf{Figure \ref{3}. Outcomes of the social dynamics.} 
  A benevolent society
  faces three possible outcomes of the social dynamics when dealing with
  an inflow of individuals from the outside: (i) benevolence is adjusted
  to a sustainable level for insiders to hold a majority in a balanced
  interplay between their tolerance and the assimilation of outsiders,
  (ii) benevolence is kept too high for too long and insiders turn to a
  minority owing to their excessive tolerance relative to the
  assimilation of outsiders, and (iii) the society is polarized due to a
  failure to adjust the benevolence of insufficiently tolerant insiders
  relative to the assimilation of outsiders. We studied cases in which
  the tolerance of insiders is unaffected (red) and affected (blue) by
  assimilated outsiders. Curves of best fit added as a visual aid.
}

Outsiders dominate when there is an imbalance between their assimilation and
the tolerance of insiders, i.e., the insiders maintain their benevolence at
a very high level or for a very long time (see the \emph{Outsider dominance}
region in Fig.~\ref{3}). If insider tolerance is insufficient relative to the
assimilation of outsiders, the society becomes polarized, and cooperation
between the two subpopulations breaks down completely \cite{Traulsen10,
Helbing10} (see the \emph{Antagonism} region in Fig.~\ref{3}). From the 
perspective of the native insider population, the two latter outcomes
are unsatisfactory. On the one hand, outsider dominance may lead to the
disappearance of the cultural patterns of insiders. On the other hand,
antagonism leads to social turmoil, thus defeating the purpose of the
initial incentive to attract outsiders.

The simplified structure of the present model makes it possible to find
a necessary condition for the state of antagonism in an analytical form.
We begin by noting that when the fraction of outsiders equals the average
tolerance, i.e., when $f_g=\langle T_r \rangle$, approximately one-half
the insiders become radicals\cite{Galam,Parravano}. This makes the fraction of radicals in the
total population $f_r=0.5 (1-f_g)$ and the fraction of tacit supporters
among the remaining insiders $f_t=0.5(1-f_g)f_g$. Thus $f_r+f_t=0.5
(1-f_g^2) < 50\%$ for any $0 < f_g \leq 1$, indicating that radicals and
their tacit supporters achieve the majority needed to engage in herd
behavior when $f_g > \langle T_r \rangle$. The condition for herd
behavior, $f_r+f_t \geq 50\%$, is thus supplemented by $f_g=(1+\epsilon)
\langle T_r \rangle$ to yield $f_r=(0.5 + \epsilon) \times (1-f_g)$,
$f_t=(0.5 - \epsilon) \times (1-f_g) f_g$, and $\epsilon \geq 0.5 f_g^2
/ (1-f_g)^2$. We find that herd behavior occurs only when the dynamics
permit $\langle T_r \rangle \leq f_g (1-f_g)^2 / (1-2 f_g+1.5 f_g^2)$
while still $f_g<50\%$. When large $p_1$ values stop the fraction of
outsiders from reaching values higher than $f_g \approx 0.2$, the
condition for herd behavior simplifies to $\langle T_r \rangle \leq f_g$.

Although the tolerance of the original insider population is preserved
in this case---i.e., outsiders who assimilate insider cultural patterns
become a part of the same distribution of tolerance---this is not always
the outcome. Assimilated outsiders may either be more tolerant towards
``their own'' than the average insider, or they may want to radically
separate from their own past. Lacking reliable data, we study a scenario
in which the tolerance of an assimilated outsider is determined using a
uniform distribution between zero and unity (see the blue curves in
Fig.~\ref{3}).  Thus tolerance turns into a dynamic variable that, depending
on the initial average tolerance, may increase or decrease the tolerance
of the society. Using the simplified condition for herd behavior,
$\langle T_r \rangle \leq f_g$, we focus on the ``race'' between the
tolerance level and the fraction of outsiders to determine whether a
society splits into two non-interacting subpopulations or ends up being
dominated by one of them. We can understand this qualitatively by
assuming that both $\langle T_r \rangle$ and $f_g$ grow over time
linearly. Thus $\langle T_r \rangle(t) = \langle T_r \rangle(0) + v_T t$
and $f_g(t) = f_g(0) + v_g t$, where $\langle T_r \rangle(0)$ and
$f_g(0)$ denote the initial values of $\langle T_r \rangle$ and $f_g$,
respectively, and $v_T$ and $v_g$ are the corresponding growth
rates. Herd behavior and a subsequent breakdown ensue when
\begin{equation}
  f_g(t) \geq \frac{\langle T_r \rangle(0) - f_g(0) }{1 -
    \frac{v_T}{v_g}} + f_g(0). 
  \label{eq2}
\end{equation}
In other words, a small initial average tolerance value and a rapid
outsider growth rate increases the probability that there will be a
breakdown.

Another scenario deserving attention is that the assimilation process may
require a minimum amount of time. Accordingly, only outsiders that entered
the system at a given moment in the past could be assimilated into insiders.
We analyzed the delay in the assimilation process by means of numerical
simulations and described the results in SI text.

\section*{Data}

\begin{wrapfigure}{l}{70mm}
\centering \includegraphics[scale=0.77]{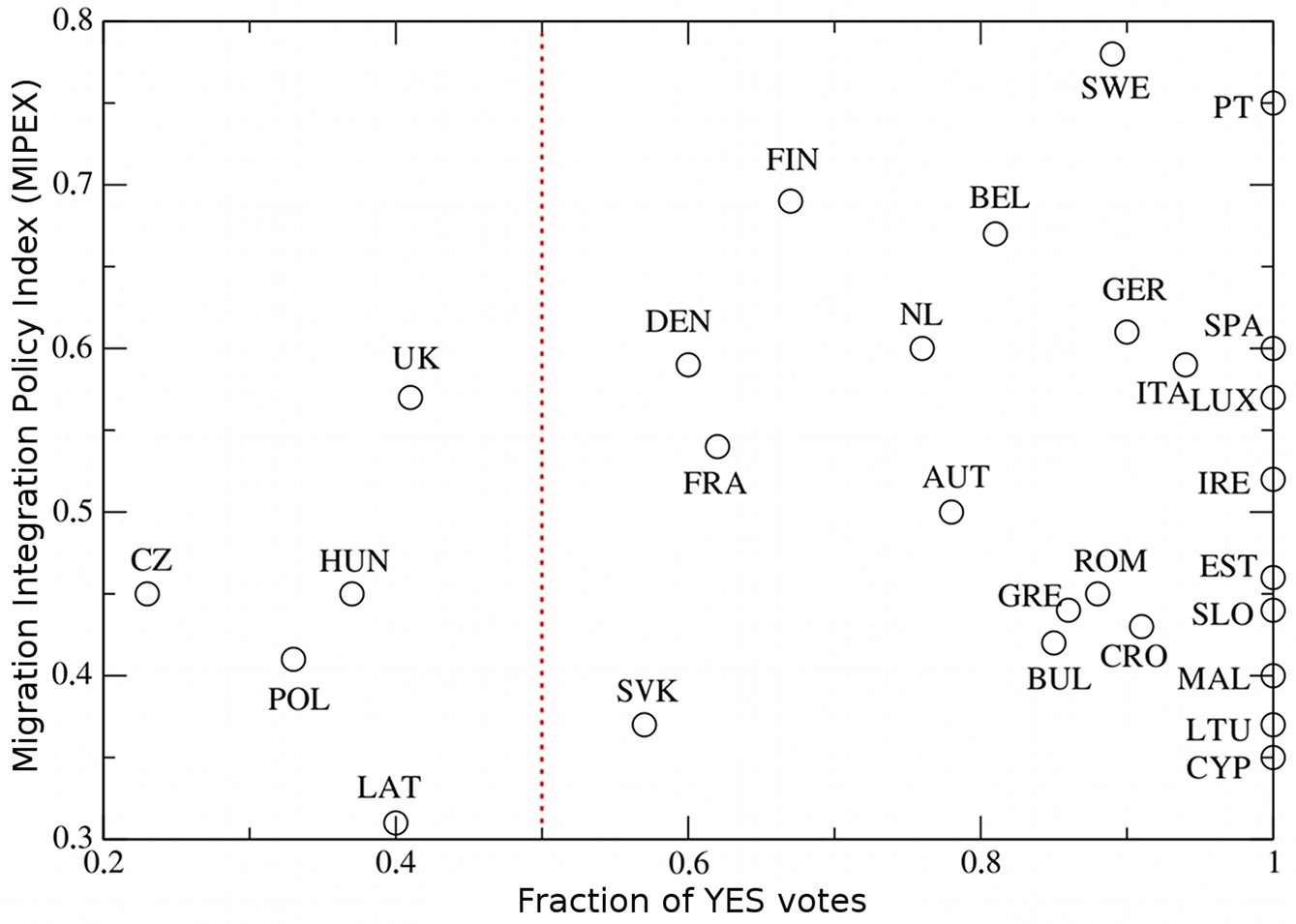}\\
\captionsetup{labelformat=empty} 
\caption{} 
\label{4} 
\end{wrapfigure}

\marginpar{
\vspace{.7cm} 
\color{Gray} 
\textbf{Figure \ref{4}. Proxies for assimilation and tolerance in the real
    world: the case of the EU countries.} 
    The net assimilation rate is quantified by MIPEX index, whereas the average tolerance is
    quantified by the fraction of pro-immigration votes. A considerable
    scatter emphasizes that countries differ in preparedness to accept
    immigrants (parameter $p_1$) and their levels of tolerance (parameters
    $T_r^{\rm min}$ and $\sigma$). Left of the red dashed line are the countries
    that are openly anti-immigrant.
}

We are using the terms ``insider'' and ``outsider'' metaphorically, but
the model can be applied to any situation with a built-in asymmetry,
e.g., if a majority provides support to a minority. When the European
Parliament recently discussed the migration of refugees into Europe,
they concluded that this current global trend is not just short-term or
temporary. When the question of whether to increase the number of
immigrants from the Middle East and sub-Saharan Africa was presented,
some EU countries voted against any additional immigration \cite{EUvote}. If we assume
that these votes are proxies for the level of a country's tolerance
towards minorities and use our model to interpret the outcome of the
voting, we find that (i) different countries have different levels of
tolerance (parameters $T_r^{\rm min}$ and $\sigma$ in the model) and
(ii) some countries have already reached a critical point in which
anti-immigrant radicals and tacit supporters are in the majority.

We extend this analogy and use the Migrant Integration Policy Index (MIPEX)
\cite{MIPEX} as a proxy for the assimilation rate (parameter $p_1$) and
plot the data in Fig.~\ref{4} using the same procedure as in Fig.~\ref{3}.
In this manner, x-axes (y-axes) of both figures point in the direction of increasing tolerance (assimilation), controlled by parameter $\sigma$ ($p_1$) in Fig.~\ref{3} and represented by the fraction of pro-immigrant votes (MIPEX) in Fig.~\ref{4}. We find that openly anti-immigrant countries end up in the left lower corner, as expected from the results of the model, but that we can obtain the EU country borderlines that delineate the different dynamical regimes only when information on the distribution of tolerance---presently approximated by a uniform distribution in the model---is available.

The lack of any clear correlation in Fig.~\ref{4} is indicative of the complexity of the real world in that at least one additional, and possibly more, explanatory factor(s) affect the relationship between assimilation and tolerance. However, despite being a very rough approximation of reality, the scatter pattern in Fig.~\ref{4} is rather informative when interpreted in conjunction with the model results.

Three groups of countries are identifiable in Fig.~\ref{4}. The countries historically exposed to large immigration inflows (e.g., the Netherlands, Portugal, Spain etc.) are both more tolerant and better equipped to receive immigrants. These countries appear in the upper-right corner of the plot, corresponding to the \emph{Mutualism} domain in Fig.~\ref{3}. By contrast, the countries in the lower-left corner (the Czech Republic, Hungary, Poland, and Latvia), corresponding to the \emph{Antagonism} domain in Fig.~\ref{3}, are former socialist states, ethnically and culturally homogeneous, and lacking substantial immigration. These countries therefore felt little pressure to enact effective immigration policies, which is reflected in their low proxy for assimilation. Furthermore, a low proxy for tolerance is less surprising if a country is unaccustomed to a large direct inflow of immigrants (e.g., Hungary) or, for that matter, to large inflows into neighboring countries (e.g., the Czech Republic and Poland in relation to Germany).

\begin{wrapfigure}{l}{70mm}
\centering \includegraphics[scale=0.77]{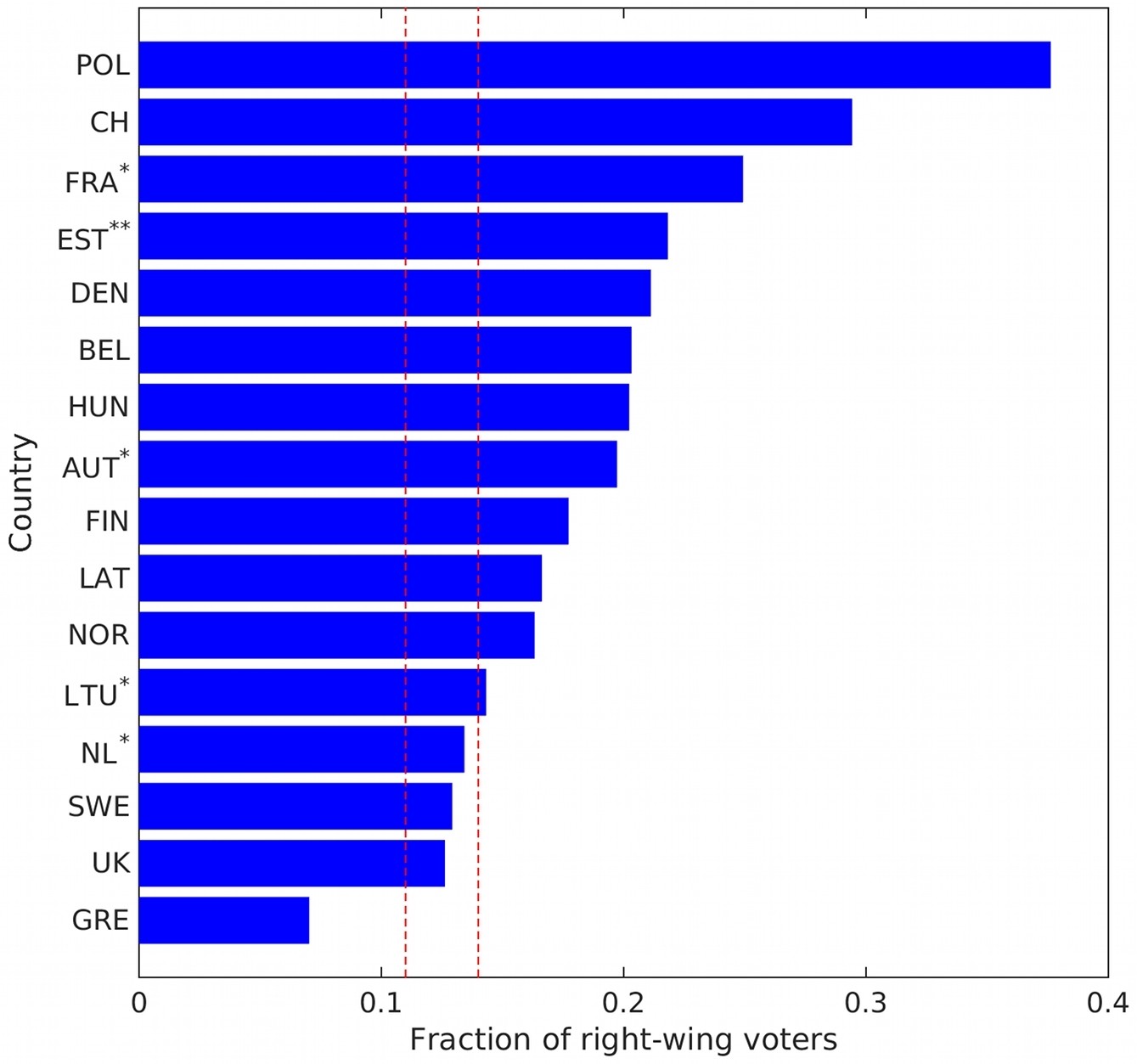}\\
\captionsetup{labelformat=empty} 
\caption{} 
\label{5} 
\end{wrapfigure}

\marginpar{
\vspace{.7cm} 
\color{Gray} 
\textbf{Figure \ref{5}. Success of the right-wing nationalist parties during the
      recent elections across Europe.} 
    Listed are the European countries that contained more than 7\% of right-wing nationalist voters in
    nationwide elections held in 2014 or 2015. The fraction of
    right-wing nationalist voters in many countries exceeded the
    percentage of the population that ``was troubled by the presence of
    people of other nationality, race, or religion'' in Europe in the
    early 1990s \cite{Betz93} (range indicated by the red dashed
    lines). Only parties that openly oppose immigration were included.
    $^{*}$For countries where national elections were held
    before 2014, we used the results of the last European parliament
    elections. $^{**}$In Estonia, the results of the two major
    right-wing nationalist parties were summed.
}

The third group of countries is found in the lower-right corner of Fig.~\ref{4}, corresponding to the \emph{Outsider dominance} domain in Fig.~\ref{3}. Among the members of this group are countries that maintained ethnic and cultural homogeneity, and/or lacked substantial immigration, which is consistent with a low proxy for assimilation. As for the high tolerance, most of these countries (not Greece though) were largely unaffected by the immigration wave prior to the EU vote because of, among other reasons, geographic locations off the main migration pathways (e.g., Estonia, Lithuania, and Ireland, but also Romania, and at the time Croatia and Slovenia). Since the vote, however, some countries were struck by the crisis more seriously and at least one (Slovenia) reversed its previous policy. The overall conclusion is that the scatter pattern in Fig.~\ref{4} is reminiscent of the domains in Fig.~\ref{3}.
 
When an incursion of outsiders decreases the level of benevolence in a
benevolent society this decrease appears to be related to the
radicalization process (see Fig.~\ref{5}). For example, in the 1980s there was
a dramatic increase in the number of refugees and illegal immigrants
entering European countries and by the early 1990s this had provoked a
wave of radical right-wing populism \cite{Betz93}. Between 11 and 14\%
of the population of Europe found other nationalities, races, or
religions unsettling, and this group became a major source of votes for
right-wing nationalist parties \cite{Betz93}. The recent migration trend
brought a new wave of success for right-wing nationalists, suggesting
that the population of many European countries has been radicalized
beyond that recorded in the early 1990s (see Fig.~\ref{5}). The fraction of
right-wing voters in a given country may not correlate with the fraction
of immigrants because (i) the EU is a supra-national entity without
internal borders between the member states, and (ii) EU countries are
not equally tolerant, i.e., each country exhibits its own value of
parameter $\sigma$. Note that supporters of right-wing nationalist
parties seem to originate from two social groups \cite{Betz93}, (i)
those in competition with non-European immigrants for jobs and housing
(identified as radicals in our model), and (ii) the ``new middle class''
consisting of highly educated, often self-employed, yet politically
restless youth who are less likely to admit belonging to the political
right (the tacit supporters in our model).

\section*{Discussion}
We have devised and analyzed a social dynamics model in which the
benevolence level of an insider population decreases in response to an
incursion of outsiders. Because this decrease is a result of
radicalization, it often generates an increase in violent incidents
\cite{Makse15, Yam07, Krueger09, Dancygier10}. This increase can function as
early-warning signal that the society is approaching a critical point in
which all relations between insiders and outsiders abruptly
terminate. The heterogeneity in the tolerance among European countries
can mean that the EU tolerance is as strong as its weakest link, and
that if the cooperation in the least tolerant EU country abruptly
terminates this breakdown can trigger a cascade of societal breakdowns
through other EU countries. This cascade phenomenon is a general
characteristic of complex systems, and thus cooperation as a
counterforce is essential if societies are to continue functioning
\cite{Traulsen10, Helbing10}. The EU is currently facing an increase in
religious intolerance and a huge inflow of immigrants, and some
societies are responding by raising walls or fences on their borders, a
strong signal that these societies have already reached a critical
point. It is thus extremely important that we understand how a society
reacts to the increasing inflow of outsiders and identify the links that
produce an increase in violent incidents. Understanding the underlying
social dynamics may help policy makers undertake legislative actions
that will prevent potential conflicts on a much larger scale. Nations
need to strike a balance between the short-term benefit of accepting
immigrants and any potential long-term violence.

One limitation of our approach is that we give the population topology a
secondary role. We assume (i) that each member of each subpopulation
(insiders and outsiders) is equally able to assimilate the cultural
patterns of the other subpopulation and (ii) that new agents who enter
the system as a part of population growth make friends randomly,
irrespective of identity tags. Because of homophily \cite{McPherson01},
we envision adding a preferential attachment (PA) factor \cite{Barabasi}
which makes new outsiders more likely to connect to older outsiders and
new insiders more likely to connect to older insiders.  Using PA changes
the topology of the population, and hubs of predominantly single
subpopulation agents appear. In this structure will the assimilation rate
be the same for, e.g., an outsider surrounded by outsiders and an outsider
in a mixed neighborhood? A more realistic treatment of the assimilation
issue under PA may be the formalism of Watts \cite{Watts01} or extension
of Watts \cite{Jusup15}, i.e., an agent may be more likely to change their
tag if they are befriended by an increasing number of agents carrying the
opposite tag. These considerations again have important implications for
a country's immigration policy. In France, for example, large immigrant
communities exist in Greater Paris, Lyon, and Marseille, suggesting that
immigration indeed tends to be a PA-type process. The same is true in
much of Western Europe, suggesting that the assimilation rate is
severely impeded by the topology of a mixed native and immigrant
population. The result is two ethnically-split subpopulations, which, in
light of the work of Esteban, Mayoral, and Ray \cite{Esteban}, is a
dangerous combination, i.e., two measures of ethnic division---
fractionalization and polarization---jointly influence an increase in
violent incidents and the possibility of intrastate wars. A potential
measure against that outcome may be exemplified by Singapore, where tenants
in government-built housing (which is 88\% of all housing) must be of mixed
ethnic origin.

Modulating the asymmetry of interactions to attain a preferred
evolutionary result is found not only in human, but in animal ``societies''
as well. For instance, a benefit differential similar to our $\Delta_b$
arises in mutualistic relationships in nature if some symbionts provide
less benefit to a host than others, and the host cannot discriminate
between them \cite{Kiers03}. The local stability of the cooperative state
may be maintained by a non-equilibrium mechanism of density-dependent
interference competition among symbionts \cite{Wang11}. However, the local
stability means that mutualism between the host and symbionts may turn into
antagonism for a period of time due to the outside (environmental)
conditions. To stabilize mutualism, hosts generally punish less cooperative
symbionts \cite{Kiers03, Clutton-Brock95}, thus pushing the system to a
cooperative steady-state. Reminiscent of Popper's statement, the host cannot
maintain unlimited tolerance towards less cooperative symbionts indefinitely.

Mathematical modeling helps us understand the relationships between the main forces shaping a phenomenon of interest. The society, however, is extremely complex in the sense that human interactions are inhomogeneous, processes occur at multiple time scales, and state variables---e.g., size, production, and wealth---are unconserved quantities. It is therefore inappropriate to treat the results of mathematical modeling as a substitute for an informed discussion between policymakers. At best, modeling results aid such discussions by making them more informed.

\section*{Supporting Information}

\setcounter{figure}{0}
\renewcommand{\thefigure}{S\arabic{figure}}
\setcounter{table}{0}
\renewcommand{\thetable}{S\arabic{table}}

\paragraph*{Equilibrium and stability.} An issue of some importance is the stability of the outcomes---mutualism, outsider dominance, and antagonism---of the social dynamics. Antagonism is a non-equilibrium absorbing state such that whether the system ends up in this state or not depends on the path from the initial condition to the equilibrium attractor(s). An analytical inequality that delineates the border between the state of antagonism and the other two states in the parameter space is presented in the main text. However, very little has been said about the model's equilibrium and stability as the key determinants for the states of mutualism and outsider dominance.

We begin by finding the necessary and sufficient condition for an equilibrium. The fraction of outsiders in the system, $f_g$, according to the model description obeys the following equation
\begin{equation}
  f_g(t+1) = \left[ R(t) - p_1(t) \right] f_g(t),
\end{equation}
where $R$ is the ratio of the fitness of outsiders to the average fitness of the whole population and $p_1$ is the assimilation rate. If the system is in an equilibrium, we have $f_g(t+1) = f_g(t)$ and therefore $p_1 = R-1$. Conversely, if the last equality is satisfied, we immediately obtain $f_g(t+1) = f_g(t)$. A necessary and sufficient condition for the equilibrium is thus $p_1 = R-1$. This straightforward condition, unfortunately, cannot guarantee that the system has only one (globally stable) equilibrium because $R$ depends on $f_g$ in a relatively complex manner despite the simplifying assumptions used in model construction.

Next, we set to find an analytical expression for the dependence of $R$ on $f_g$ in hope to learn more about the model's equilibrium and the overall dynamics. To that end, it is first necessary to examine the interactions between agents in more detail. We said that at each time step, $m$ pairs of agents were randomly selected to cooperate with each other, where the result of this cooperation depended on who played the role of a donor and a recipient. When selecting a random pair of agents, there are four possible outcomes, which we denote $(I,O)$, $(I,I)$, $(O,O)$, and $(O,I)$, with $O$ ($I$) indicating an outsider (insider). Let us also denote with $p_i$, $i \in \lbrace 1,..., 4 \rbrace$ the corresponding probabilities.

To find these probabilities, we note that in the case of, say, $p_1$, an insider is randomly selected with the probability determined by the abundance of insiders, i.e., $1-f_g$. However, we also need to know the probability of randomly picking an outsider from the insider's neighborhood. Among the connections an insider has at any moment in time, there are $\langle k \rangle f_g X(f_g)$ outsiders and $\langle k \rangle (1-f_g)$ insiders, where $\langle k \rangle$ is the average degree at the beginning and $X(f_g)$ is the probability that the insider has not severed its links with outsiders. Because we assume a uniform distribution of tolerance, $\mathcal{U}(T_r^{\rm min}, T_r^{\rm min}+\sigma)$, it turns out that $X(f_g) = 1 - \min \lbrace \max \lbrace 0, U(f_g) \rbrace, 1 \rbrace$, where $U(f_g) = (f_g - T_r^{\rm min}) / (\sigma - T_r^{\rm min})$. By taking the ratio between the number of the insider's connections to outsiders and the total number of connections that this insider has, we get the desired probability (of randomly picking an outsider in the insider's neighborhood), $f_g X(f_g) / (f_g X(f_g) + (1-f_g))$. It follows that
\begin{equation}
  p_1 = (1-f_g) \times \frac{f_g X(f_g)}{f_g X(f_g) + (1-f_g)}.
\end{equation}
Analogous reasoning leads to the following expressions
\begin{eqnarray}
  p_2 &=& (1-f_g) \times \frac{1-f_g}{f_g X(f_g) + (1-f_g)}, \\
  p_3 &=& f_g \times \frac{f_g}{f_g X(f_g) + (1-f_g)}, \mathrm{and} \\
  p_4 &=& f_g \times \frac{(1-f_g) X(f_g)}{f_g X(f_g) + (1-f_g)}.
\end{eqnarray}

\begin{table}[!t]
  \centering
  \caption{\small {\bf Definitions of the interactions between the different pairs of agents.}}
  \begin{tabular*}{\hsize}{@{\extracolsep{\fill}}cccc}
    Pair               & Probability & Fitness         & Fitness         \cr
    (recipient, donor) & $p_i$       & $\Phi_1$        & $\Phi_2$        \cr
    \hline
    $(I,O)$            & $p_1$       & $+b$            & $-c$            \cr
    $(I,I)$            & $p_2$       & $+(b-c)$        & 0               \cr
    $(O,O)$            & $p_3$       & 0               & $+(b-c)$        \cr
    $(O,I)$            & $p_4$       & $-(c+\Delta_c)$ & $+(b+\Delta_b)$ \cr
    \hline
    $b$ and $c$, $b>c$        & \multicolumn{3}{l}{benefit and cost of cooperation, respectively.} \cr
    $\Delta_b$ and $\Delta_c$ & \multicolumn{3}{l}{benefit and cost differentials, respectively.}  \cr
    \hline
  \end{tabular*}
  \label{tS1}
\end{table}

Using probabilities $p_i$, $i \in \lbrace 1,..., 4 \rbrace$, it is possible to analytically express the fitness of insider, $\Phi_1$, and outsider, $\Phi_2$, subpopulations. Fitness is defined as the average per-capita benefit net of the cost of cooperation. A convenient way to calculate fitness is (i) to separately find the average net benefit resulting from the interactions that involve insiders on the one hand and outsiders on the other hand, and (ii) to divide these quantities with the average number of insiders and outsiders, respectively, per interaction. For example, if $(I,O)$ pair interacts, then we have one insider and one outsider taking part in a single interaction, but if the pair is $(I,I)$, then a single interaction involves two insiders, which has to be taken into account. How fitness is affected when any of the four possible pairs of agents interact is presented in Table~\ref{tS1}.

Before writing down the expressions for fitness, we note that the setup in Table~\ref{tS1} is dependent only on the magnitude of costs relative to benefits. Accordingly, in all numerical simulations, $c/b = \Delta_c/\Delta_b = 10\%$, with $\Delta_b/b = 25\%$. First we address the fitness of the insider subpopulation, calculated here as the aforementioned division of (i) the average net benefit from the interactions involving insiders and (ii) the average number of insiders per interaction. The equation is
\begin{equation}
  \Phi_1 = \frac{0.5 (p_1+p_4) b + p_2 (b-c) - 0.5 (p_1+p_4) (c+\Delta_c)}{p_1 + 2 p_2 + p_4}.
  \label{eqS6}
\end{equation}
The numerator in this equation is practically read off from Table~\ref{tS1}, with one twist---after randomly drawing pairs $(I,O)$ or $(O,I)$, who acts as a donor and who acts as a recipient is still undecided, yet each role is equally likely, thus causing the term $0.5 (p_1+p_4)$ to appear. The denominator expresses the idea that pairs $(I,O)$ and $(O,I)$ appear with probabilities $p_1$ and $p_4$, respectively, and contain a single insider, whereas pair $(I,I)$ appears with probability $p_2$, but contains two insiders. An analogous equation for the fitness of outsiders is
\begin{equation}
  \Phi_2 = \frac{-0.5 (p_1+p_4) c + p_3 (b-c) + 0.5 (p_1+p_4) (b+\Delta_b)}{p_1 + 2 p_3 + p_4}.
  \label{eqS7}
\end{equation}

One last step in the present analysis is to combine Eqs.~[\ref{eqS6}] and [\ref{eqS7}] into the fitness ratio, $R=\Phi_2/\left[ \Phi_1 (1-f_g) + \Phi_2 f_g \right]$. Function $R$ defined in this way faithfully reproduces the curves in Fig.~\ref{1} obtained by means of numerical simulations. More importantly, in a typical model setup whereby outsiders are attracted into the system, function R monotonically decreases until an equilibrium point is reached (unless the model trajectory gets trapped beforehand in the absorbing state, i.e., the state of antagonism). Because $R$ is monotonically decreasing, there can be only one point that satisfies condition $p_1 = R-1$, thus indicating that the model equilibrium is globally stable.

\begin{wrapfigure}{l}{70mm}
  \centering \includegraphics[scale=0.77]{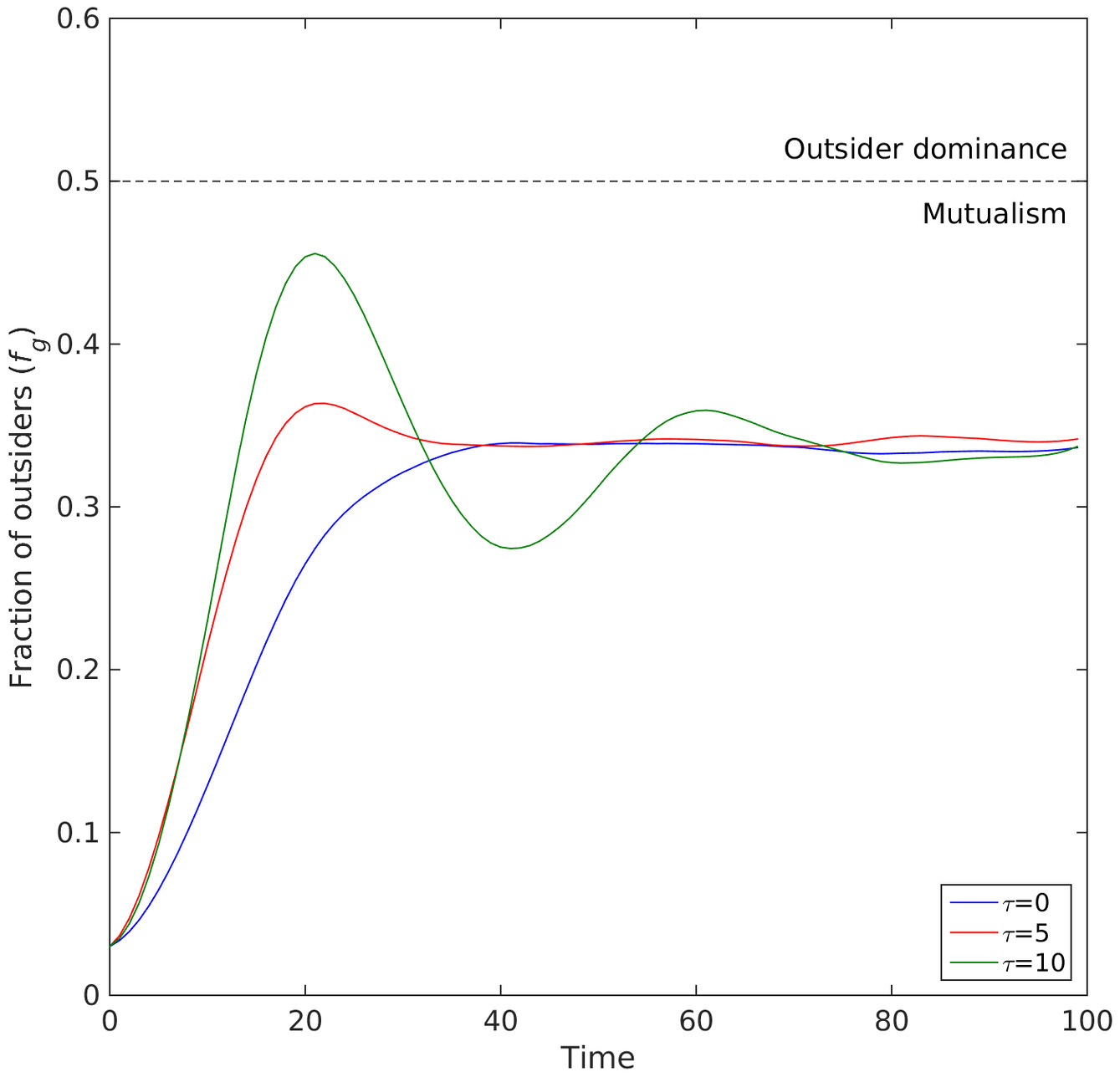}\\
\captionsetup{labelformat=empty} 
\caption{} 
\label{S1} 
\end{wrapfigure}

\marginpar{
\vspace{.7cm} 
\color{Gray} 
\textbf{Figure \ref{S1}. Delayed assimilation excites damped oscillations without shifting the model's equilibrium point.} 
An increasing delay in the assimilation process leaves the equilibrium point of the model unchanged, yet modifies the dynamics in two important ways: (i) as the fraction of outsiders starts to increase, delayed assimilation affects less individuals than instantaneous assimilation, thus allowing a sharper rise of outsiders in the system and (ii) because the assimilation process trails the current state of the system, the fraction of outsiders overshoots the equilibrium point and sets off damped oscillations. The longer the delay, the higher the amplitude of oscillations. The average tolerance is $\langle T_r \rangle = 0.40$. The assimilation rate is $p_1=0.1$.
}

\paragraph*{Delayed assimilation.} In the main text, we assumed that the assimilation process is effective instantaneously, whereas in reality it may take some time for outsiders to assimilate new cultural patterns. This possibility is readily taken into account by introducing delay $\tau$, such that the fraction of outsiders present in the system at time $t-\tau$ is assimilated at rate $p_1$. Here, we examine how introducing delayed assimilation affects the model dynamics and discuss several important implications of the obtained results.

Delay in the assimilation process brings about two noticeable changes in the model dynamics (Fig.~\ref{S1}). First, if the fraction of outsiders is increasing, fewer individuals are influenced by delayed assimilation in comparison to the instantaneous assimilation. A consequence is that the fraction of outsiders increases at a higher rate when there is some delay than when there is none. Second, delayed assimilation leaves the model equilibrium unaffected, but---by trailing the current state of the system---makes it possible that the fraction of outsiders temporarily overshoots the equilibrium point and sets off damped oscillations. These new features in the model dynamics change the critical average tolerance that marks the border between a mutualistic and an antagonistic relationship.

\begin{wrapfigure}{l}{70mm}
  \centering \includegraphics[scale=0.77]{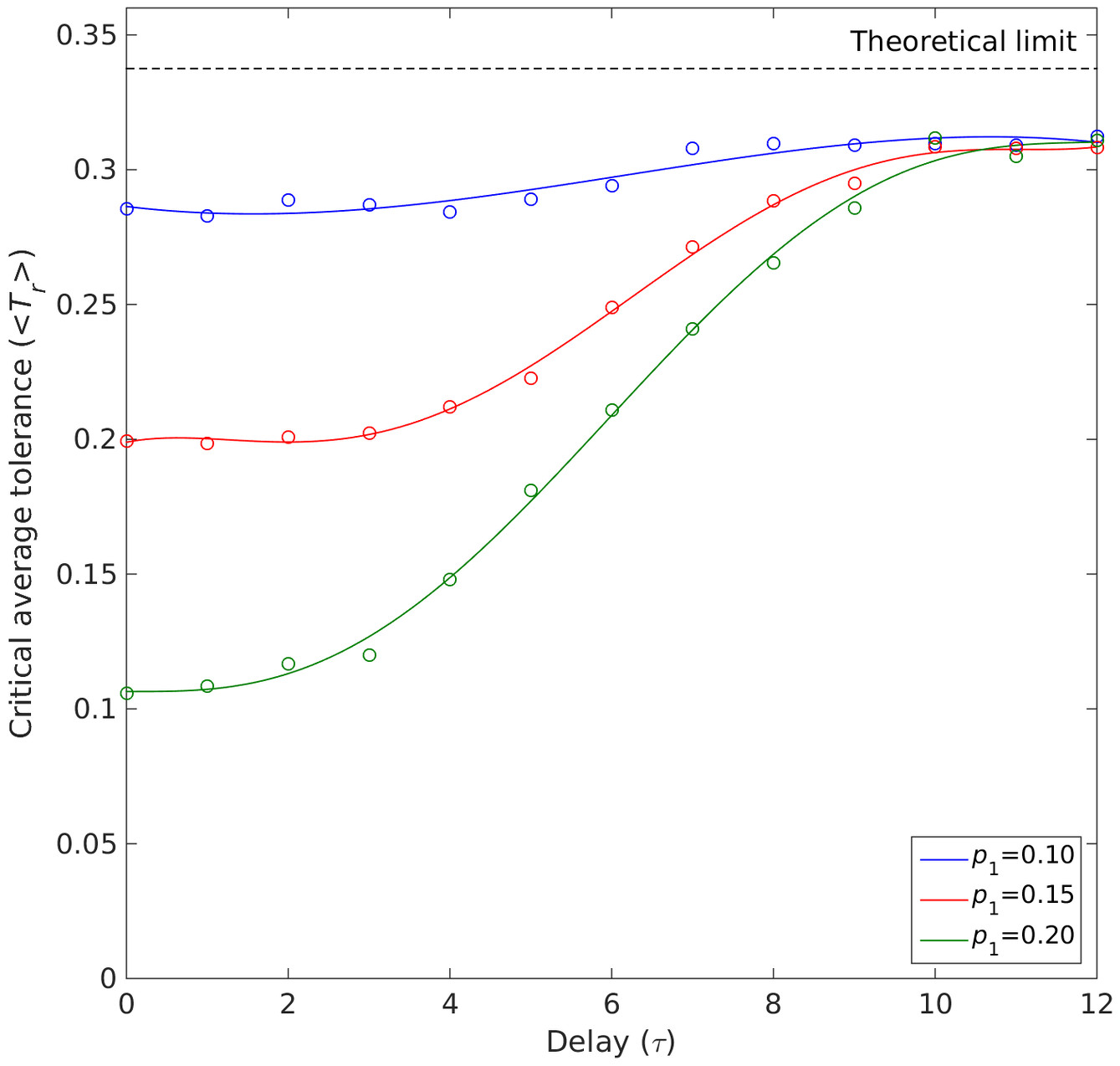}\\
\captionsetup{labelformat=empty} 
\caption{} 
\label{S2} 
\end{wrapfigure}

To reach mutualism under delayed assimilation, a society must be more tolerant than under instantaneous assimilation (Fig.~\ref{S2}). Namely, the temporary accumulation of outsiders above the equilibrium point, which is impossible without a delay, may just be enough to push an otherwise mutualistic system into antagonism. Because antagonism is a non-equilibrium absorbing state, it is irrelevant that the delay has no effect on the model's equilibrium. What matters is the path that leads to the equilibrium point, and that path turns less favorable for mutualism with the introduction of delayed assimilation. The higher the delay, the less likely the mutualistic relationship. Accordingly, the critical average tolerance increases with the delay ($\tau$) from a lower limit at $\tau = 0$ to a theoretical upper limit (see the main text) beyond which the state of antagonism disappears altogether. This lower limit decreases with the assimilation rate ($p_1$), yet the theoretical upper limit is independent of $p_1$. Consequences are that (i) the described effect of delayed assimilation is stronger at high values of $p_1$ and (ii) the critical average tolerance becomes independent of $p_1$ at high enough $\tau$.

\begin{figure}[h]
  \centering \includegraphics[scale=0.77]{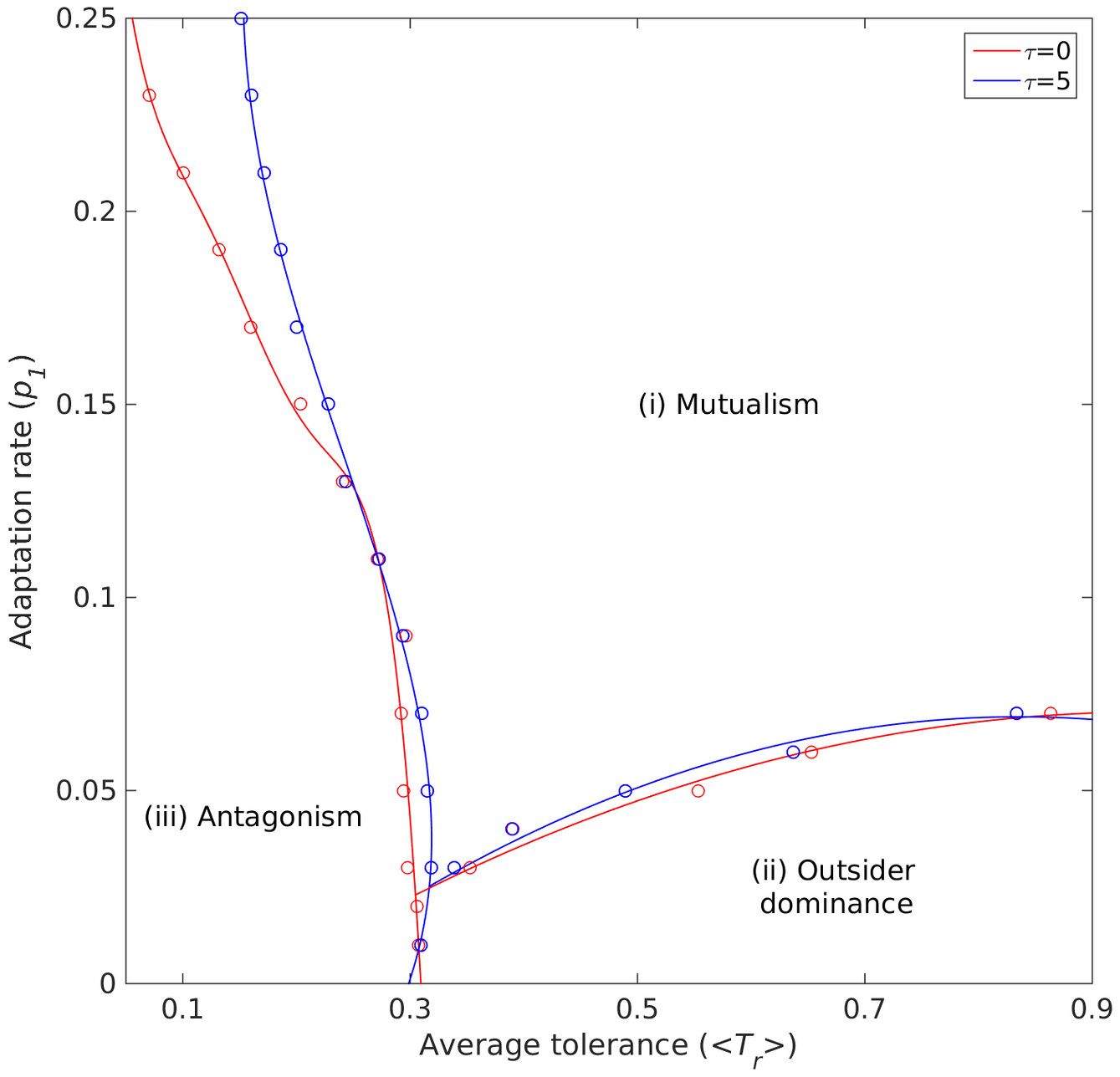}\\
  \caption{{\bf Outcomes of the social dynamics with delayed assimilation.} In much of the parameter space, delayed assimilation has a minor effect on the mutualism of insiders and outsiders, except when the rate at which outsiders are assimilated is relatively high. At high assimilation rates, mutualism gives way to antagonism because the delay causes the system to temporarily accumulate more outsiders  than it would be possible otherwise, which in turn triggers herd behavior even among more tolerant insiders.}
  \label{S3}
\end{figure}

How are the overall outcomes of the social dynamics changed by delayed assimilation? To answer this question, in Fig.~\ref{S3}, we provide a map of the parameter space under instantaneous assimilation ($\tau = 0$) overlaid with the same kind of map under delayed assimilation ($\tau = 5$). As expected from point (i) above, there is very little change in the results if the assimilation rate is relatively low. The results, by contrast, change markedly at high assimilation rates, with mutualism giving way to antagonism. We notice that in line with point (ii) above, the border delineating the states of mutualism and antagonism exhibits less dependence on the assimilation rate ($p_1$) than in the case of instantaneous assimilation. Having longer delays than $\tau = 5$ (see Fig.~\ref{S2}) would erase this dependence altogether because the border between mutualism and antagonism would be pushed toward its theoretical limit, which is independent of $p_1$.

\marginpar{
\vspace{.7cm} 
\color{Gray} 
\textbf{Figure \ref{S2}. Society must be more tolerant for delayed assimilation to be effective.} 
Critical tolerance level marking the border between mutualism and antagonism increases with the delay in the assimilation process. This increase is a consequence of the temporary accumulation of outsiders above the equilibrium point, which is impossible without a delay. Note that the critical average tolerance has a lower limit dependent on the assimilation rate ($p_1$) and a theoretical upper limit (see the main text) denoted by a dashed line that is independent of $p_1$. Consequently, (i) the effect shown in this figure strengthens with $p_1$ and (ii) the critical average tolerance becomes independent of $p_1$ at high enough $\tau$.
}

The outcomes of the social dynamics under delayed assimilation have an important implication in the context of immigration policies. It turns out that having in place very efficient assimilation programs (corresponding to a high value of $p_1$) may not mean they are effective. To achieve the effectiveness, it is also necessary that such programs produce the expected results rather quickly (corresponding to a low value of $\tau$). An alternative path to avoiding excessive radicalization is to introduce measures that improve the average tolerance of insiders. Provided $p_1$ is high enough, a higher average tolerance favors the prospect of reaching a mutualistic equilibrium even if there is a considerable delay. An inevitable conclusion is that a successful immigration policy is a tough balancing act that requires people to make concessions in order to learn how to live together.

\section*{Acknowledgments}
We are grateful to Robert Axelrod, Robin Dunbar, Yoh Iwasa, J{\"u}rgen Kurths, and Tomislav Lipi{\'c} for helpful suggestions. B.P. and H.E.S. received support from the National Science Foundation (NSF) Grant CMMI 1125290. B.P. also received support from the University of Rijeka. M.J. was partly supported by the Japan Science and Technology Agency (JST) Program to Disseminate Tenure Tracking System. Z.W. was supported by the National Natural Science Foundation of China, Grant No. 61201321 and 61471300.

\nolinenumbers

\end{document}